# An unexpected role of H during oxidation of SiC in water

Jianqi Xi[1], Cheng Liu[1], Dane Morgan[1], Izabela Szlufarska[1,2*]

[1.] Department of Materials Science and Engineering, University of Wisconsin, Madison, WI, 53706, USA.
[2.] Department of Engineering Physics, University of Wisconsin, Madison, WI, 53706, USA

**Abstract**

During aqueous oxidation, atoms in the solid react chemically with oxygen, leading either to formation of an oxide film or to dissolution of the host material. Commonly, the first step in oxidation involves oxygen atom from dissociated water reacting with surface atoms and breaking near-surface bonds. In contrast, hydrogen on the surface often functions as a passivating species. Here, we discovered that the roles of O and H are reversed in the early oxidation stages on SiC. O forms stable species on the surface and chemical attack occurs by H breaking Si-C bonds. This so-called hydrogen scission reaction is enabled by a newly discovered metastable bridging hydroxyl group that can form during water dissociation. Si atom displaced from the surface during water attack subsequently forms $SiO(OH)_2$, which is a known precursor to formation of silica and of silicic acid. This study suggests that the roles of H and O in oxidation need to be reconsidered.

**Introduction**

Oxidation of materials in water is of importance in numerous technological applications, including processing of semiconductor devices, cladding in nuclear reactors, and structural materials in navy applications[1–5]. Not surprisingly, aqueous oxidation has been a subject of many experimental and theoretical studies. In general, understanding of oxidation of materials in water can be divided into three steps: (i) adsorption of water molecules onto the surface, (ii) breaking of near-surface bonds, and (iii) surface dissolution or oxide growth. In pure water, near-surface bonds are typically broken due to a chemical attack by an oxygen atom formed during water dissociation on the surface. As a representative example for metal alloys, Das *et al.* investigated early stages in oxidation of Ni-Cr alloys in water and found that interaction of oxygen from $H_2O$ molecules with surface metallic atoms was responsible for surface oxidation[6]. In this process, the dissociated hydrogen atoms either passivate the surface or diffuse inside the bulk and are trapped as interstitials [6]. Similar oxidation mechanisms have also been observed in the growth of functional oxide films on semiconductors[7,8]. In particular, Weldon *et al.* investigated oxidation mechanisms of silicon in water via a combination of surface infrared absorption spectroscopy (IRAS) and density functional theory (DFT) calculations[7]. They found that oxygen atoms, transferred from the dissociated hydroxyl groups, insert themselves into the Si-Si backbonds (the chemical bond between the top surface atom and the subsurface atom), leading to formation of an electrically insulating layer. Similarly as on metal surfaces, hydrogen atoms on Si were found to passivate the surface states. These and similar observations suggest that oxygen atoms formed during $H_2O$ dissociation in general act as oxidants, whereas hydrogen atoms play the role of anti-oxidizing species that passivate the dangling bonds on intrinsic semiconductor surfaces.

---

[*] *Department of Materials Science and Engineering, University of Wisconsin, Madison, WI, 53706, USA.*
*E-mail: szlufarska@wisc.edu; Tel: +1-608-265-5878*

Silicon carbide is a versatile material that has already been used, and is being considered for new uses, both in structural and electronic applications. In the semiconductor industry, wet oxidation of SiC substrate is used to create a silica ($SiO_2$) film, which is an insulator and acts as a passivation layer for electronic devices[9,10]. In the nuclear energy field, hydrothermal oxidation of SiC is considered a critical degradation mechanism during use of SiC as a cladding material[11–13]. While it is agreed that during dry oxidation SiC forms a passive silica film, upon exposure to high-temperature water SiC has been reported to dissolve without forming a protective layer[12,13]. It has been hypothesized that surface dissolution mechanism of SiC in water could be either via direct formation of soluble silicic acids, such as $Si(OH)_4$[11,14], or via formation of a silica film that rapidly dissolves in water with a small activation energy barrier[13]. In either case, the kinetic mechanisms and rate-limiting steps for surface oxidation underlying and controlling SiC dissolution are not well established.

Considerable progress has been already made in understanding of the initial adsorption of water molecules on SiC surface[15–21]. Most of these studies focused on elucidating the details of $H_2O$ chemisorption onto the surface. For instance, IRAS experiments on the Si-rich (3×2) reconstructed SiC(001) surface found that the dissociation of water results in the passivation of the outermost excess Si layer via hydride (Si-H) and hydroxyl (Si-OH) groups [19,20]. These results are consistent with *ab initio* molecular dynamic (AIMD) simulations on *p*(2×1) reconstructed SiC(001) surfaces, which showed that dissociation of water molecule forms ordered passivating Si-H and Si-OH species[15,18]. In addition, using density functional theory (DFT) calculations, Cicero *et al.* predicted existence of bridged siloxane (Si-O-Si) groups on the Si terminated SiC(001) surface, which is formed by oxygen atom first increasing the Si-Si dimer distance and then breaking the dimer bond[16]. This finding was confirmed by the IRAS spectra collected after wet oxidation of Si at 673 K[21]. These and similar studies identified in general three types of possible water dissociation products formed during adsorption onto SiC surface, which are hydride, hydroxyl, and siloxane groups. However, chemical attack of these species on SiC during hydrothermal oxidation, beyond the initial water dissociation, have not been identified. In this study we carry out atomistic simulations of SiC surface in contact with high-temperature water in order to understand the mechanisms of hydrothermal oxidation, i.e., revealing how water attacks Si-C bonds and how the products of dissociation reactions are incorporated into the surface.

Here, we start our study with Si-terminated *p*(2x1) SiC(001) as a representative surface, because it is known to be stable[22] and it is consistent with the earlier *ab initio* calculations of water chemisorption[15–18]. Possible reactions on a surface are difficult to guess and therefore we used high temperature classical molecular dynamics (MD) simulations, which are carried out using reactive force field ReaxFF[23], to identify potential reaction candidates. Once the reactions were identified, we used DFT to determine reaction energies and activation energy barriers.

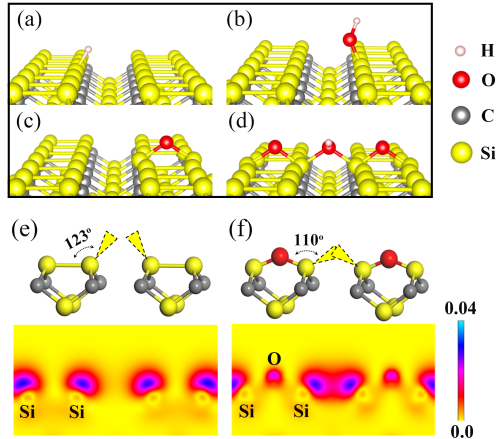

*Fig. 1. (a)-(d) Schematic structure of basic species after water dissociation, and (e)-(f) charge density distribution for the formation of siloxane. (a) hydride, (b) hydroxyl group, (c) siloxane group, and (d) bridging hydroxyl group. Charge density distribution before (e) and after (f) the formation of siloxanes. The unit for the legend bar is $e/Å^3$. Si atoms are yellow, C atoms are gray, O atoms are red and H atoms are white.*

## Methods

Our classical MD simulations were carried out using reactive force field ReaxFF[23], as implemented in Large-scale Atomic/Molecular Massively Parallel Simulator (LAMMPS) package[24]. SiC surface was created by removing periodic boundary conditions along the [001] direction. The sample contained 42 atomic layers, 64 atoms per layer, and the vacuum region above the surface was approximately 40 Å thick. Reactions were studied on the top surface, while the atoms near the bottom surface (i.e., within 10 Å-thick layer) were not allowed to relax. To prepare a model of water environment, we first created a cell (the cell size is the same as the vacuum region) containing 100 water molecules. Positions of water molecules were random and were first chosen using Packmol code[25]. Subsequently, we relaxed the system for 100 ps at 1000 K with the time step of 0.25 fs and using periodic boundary conditions in all spatial directions. Temperature was controlled using Nose-Hoover thermostat. After relaxation, we inserted these water molecules into the vacuum region above SiC surface, and then performed simulations at 1000 K to simulate the hydrothermal oxidation process for 1 ns.

DFT calculations were carried out using the Vienna Ab Initio Simulation Package (VASP)[26]. The projected augmented wave (PAW) potentials[27] were used to mimic the ionic cores, and the generalized gradient approximation (GGA)[28] method was applied for the exchange and correlation functional. In each DFT supercell, there was a slab of seven atomic layers (32 atoms per layer) for SiC. We used a vacuum layer 15 Å or greater to avoid interactions between neighboring slabs. Adsorption calculations were also performed on 13 layers and 16 atoms per layer with same vacuum region, to check convergence of values with the respect to number of slab layers. It was found that the difference in adsorption energies for species (Si-OH and Si-O-Si) on both slabs was smaller than 50 meV/adsorbate atom; therefore, we confirmed that the slab with seven atomic layers is thick enough. The lower three atomic layers were fixed in the bulk configuration with the PBE predicted bulk SiC lattice constant of 4.378 Å [29]. The plane wave cutoff energy was 500 eV, and spin-polarization and dispersion effects (described below) were considered in the calculations. The Brillouin Zone integration was performed by using the 3×3×1 Monkhorst-Pack $k$-point sampling. Relaxation was carried out until forces on all ions were lower than 0.01 eV/ Å. Solvent effects have been considered by using an implicit Poisson-Boltzmann solvent model[30]. The dielectric constant ε of water is taken to be approximately 20 for hydrothermal conditions[31]. Zero-point energy and entropy corrections were determined assuming that the adsorbates are bonded strongly enough to have negligible translational and rotational motion. The details were explained in Supplementary Materials. Dispersion interactions have been taken into account using semi-empirical corrections (vdW-D), as proposed by Grimme *et al*.[32]. Reaction energy barriers were calculated using the climbing-image nudged elastic band method[33]. Linear interpolation was used to generate 5 images for optimization. We assumed that the solvent effect on reaction barriers is negligible, which is reasonable when considering the small water dielectric constant at hydrothermal condition. Indeed, the solvent effect on the reaction energy was estimated to be less than 0.10 eV, based on the implicit solvent model, which further verifies our assumption. After hydrogen scission reaction (discussed in the results section), additional *ab initio* MD simulations were performed to identify subsequent steps in the oxidation process. Simulations were performed at 1000 K for 100 ps, and the settings of AIMD simulations were discussed elsewhere[22].

**Results and Discussion**

Based on classical MD simulations, we identified three types of basic species that form on SiC after water dissociation (see Fig. 1), i.e., hydride (Fig. 1(a)), hydroxyl (Fig. 1(b)), and siloxane (Fig. 1(c)). These species are the same as reported in previous studies[15–18]. The DFT optimized distances for Si-H and Si-O (in the Si-OH) are ~1.47 Å and ~1.66 Å, respectively. In the siloxane, two Si atoms that originally formed part of a dimer surface reconstruction nearly recover their bulk positions in a tetrahedral coordination so that the remaining dangling bonds on these atoms are tilted toward the surface and tend to lie flat, as schematically shown in Figs. 1(e) and 1(f). The optimized Si-O bond length is ~1.70 Å, and the Si-O-Si bond angle is ~123.5 °, in agreement with published calculations[16]. We found that siloxane groups are formed by transition reactions from hydroxyl groups (i.e., Si-OH + *Si → Si-O-Si + Si-H, here * denotes a site on the surface); this finding is consistent with previous experimental reports[20]. Enthalpies for these reactions depend on whether neighboring sites on SiC surface are occupied and they vary from -1.39 to -2.19 eV. The corresponding free energies at 1000 K range from -1.23 to -1.93 eV. This result suggests that termination with siloxane groups is more energetically favorable than with hydroxyl groups. Activation barriers for the aforementioned transition reactions are found to be between 0.15 and 0.64 eV. The low activation barrier indicates that the hydroxyl groups on the surface are metastable and readily transfer to siloxane groups, especially at high temperature. As a result, it is expected that a stable surface is likely to be covered with a high density of siloxanes.

Although the hydride, hydroxyl, and siloxane products of water dissociation on SiC are known, it remains to be determined how these basic species attack SiC surface bonds and enable subsequent oxidation. It is known that during dry oxidation of SiC, oxygen gas molecules dissociate on the surface and break surface bonds to form oxide films[34]. It is possible that in the water environment, oxygen atoms from $H_2O$ molecules dissociation might attack the surface as well. In fact, oxygen attack from water has been reported in earlier investigations of wet oxidation of Si[7,8]. Specifically, it is generally accepted that on Si surface, an oxygen atom transferred from -OH group, inserts itself into the Si-Si backbond. We have calculated DFT energies of analogous reactions on SiC where oxygen atom is inserted into the Si-C backbond and surprisingly we found such reactions to be energetically unfavorable (Supplementary Fig. S2).

The mechanisms of attack might be difficult to guess *a priori* and therefore we have turned to high-temperature classical MD simulations for insights. An important finding from these simulations is the presence of a fourth water dissociation product, which is a bridging hydroxyl group shown in Fig. 1(d). This bridging hydroxyl can form by transforming from hydroxyl groups that had previously formed during water dissociation according to the following reaction: Si-OH + *Si → Si-OH-Si. Interestingly, DFT calculations show that such forward reaction in isolation is energetically unfavorable (Si-OH-Si group spontaneously reverses toward Si-OH group), which explains why it has not been reported in earlier DFT studies of chemisorption of isolated water molecules. However, this reaction becomes favorable (in fact it is a barrierless transition) when siloxane bridges are present on neighboring surface sites. The presence of neighboring siloxane bridges is expected given that siloxane bridge is the most stable water dissociation product on SiC surface.

The basic process for forming a bridging hydroxyl can be explained as follows. When two bridging siloxanes form on neighboring sites (see Fig. 1(f)), the closely spaced dangling bonds from two neighboring Si dimers interact to share electrons. Consequently, an -OH group binding to one of these Si atoms will immediately interact with the second Si atom, forming a stable Si-OH-Si complex. The optimized -OH group resides exactly in the middle between the two Si atoms and creates two weak Si-O bonds. Similarly as in the case of an isolated hydroxyl group, H in the -OH group transfers charges to the oxygen in the bridging hydroxyl group. The O-H bond length in the bridging hydroxyl group is ~1.00 Å, which is slightly longer than that in the isolated hydroxyl group, ~0.97 Å. The presence of the O-H bond reduces charge transfer in the Si-O bonds and results in a longer Si-O bond (~1.93 Å) in the case of bridging hydroxyl group than in the case of bridging siloxane group where the Si-O bond is ~1.70 Å. Similar bridging structures have been observed in oxides, such as quartz-$SiO_2$ and $\theta$-$Al_2O_3$ etc., where the H interstitial atom loses electron to oxygen and forms the bridging -OH group[35,36]. As mentioned above, formation of a bridging hydroxyl is due to a shorter distance between two Si atoms that participate in neighboring siloxane bridges on the surface (see Fig. 1(f)) and due to the resulting interaction of dangling bonds. Specifically, when the two Si atoms participate in siloxane bridges, the distance between them decreased to 3.34 Å from the distance of 3.74 Å corresponding to unoxidized surface.

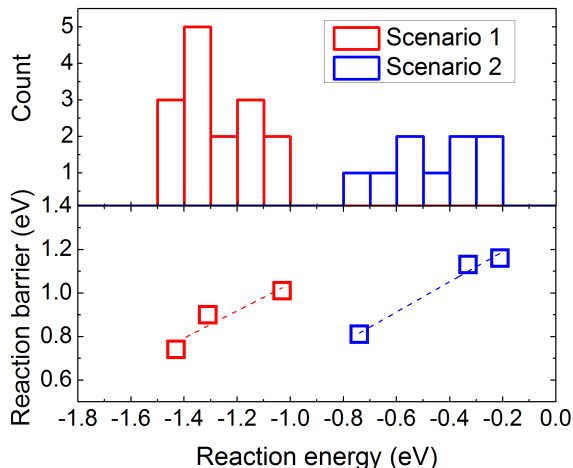

*Fig. 2. Distribution of reaction enthalpies and the corresponding activation energies for the hydrogen scission reaction that starts from a bridging hydroxyl group. Scenarios 1 and 2 correspond to cases without and with pre-existing siloxanes near reaction pathway, respectively.*

The newly found hydroxyl bridge species turns out to play a key role in surface attack as it serves as an intermediate species in the attack of water on Si-C bonds. Specifically, we found that the hydroxyl bridge is metastable and it can transform following the reaction: Si-OH-Si + (Si-C)$^b$ → Si-O-Si + *Si + C-H, where (Si-C)$^b$ stands for the Si-C backbond on the surface. H released from the -OH bridge breaks the Si-C backbond and disorders the surface. This is a consequence of the electrostatic interaction between subsurface C layer and the hydrogen atom in the -OH group, which competes with the interaction between O and H in the Si-OH-Si complex. The reaction energy depends on the termination of neighboring sites. We have considered 24 independent configurations with respect to the different neighbors around the active site (Supplementary Fig. S8). The enthalpies for the reaction in the presence of these configurations range from -0.21 to -1.47 eV, as shown in Fig. 2 (the entropic contributions to the reaction are estimated to be less than 0.08 eV even at 1000 K, Supplementary Fig. S9). Although the surrounding environment affects the reaction energy, the change in the activation energy is relatively modest (the average activation energy is ~0.96 eV and it ranges from 0.74 to 1.16 eV). The relatively small activation energy suggests that hydrogen atoms are likely to insert themselves into the Si-C bonds through a hydrogen scission reaction. After this reaction, H atom occupies the Si sublattice and forms a stable C-H bonds. The kicked-out Si atom is displaced above the surface (by at least 0.5 Å) and is exposed to the water environment. This Si atom subsequently binds either to the neighboring surface Si atom (Reaction 1) or to the adsorbed O atom (Reaction 2), as shown in Table 1. The Bader charge analysis shows that in both scenarios, the displaced Si atoms have excess electrons (~0.65|e|, see Supplementary Fig. S3), resulting in a temporary increase of the surface reactivity after hydrogen scission reactions. As a consequence, these Si atoms would easily bind to the oxidant species, such as -OH *etc.*, from $H_2O$.

H scission reaction from intermediate bridging hydroxyl groups is an important step in the attack of water on Si-C bonds because H scission reactions from other surface species are energetically unfavorable. H scission reactions from all the stable species (products of $H_2O$ dissociation) are listed in Table 1. Reactions 3 and 4 correspond to the hydrogen scission from the hydride group without and with the presence of additional siloxane groups on neighboring sites, respectively. Both cases involve significant energy barriers (at least 2.51 and 2.18 eV, respectively)-consistent with the observations that Si-H bonds are generally very stable[16]. Reaction 5 corresponds to the hydrogen scission from a hydroxyl group. In this reaction, H atom is released from the isolated hydroxyl group and it breaks the Si-C bond. The remaining oxygen atom forms a siloxane bridge by bonding an additional Si atom. The product of this reaction is similar to that in the bridging hydroxyl group (Reaction 1 and 2). However, the energy barrier for the reaction is higher in the case of H scission from a hydroxyl group (Reaction 5) than from hydroxyl bridges (Reactions 1 and 2), due to the stronger O-H bond in an isolated hydroxyl group than that in the bridging hydroxyl group[16]. These results suggest that the hydrogen scission reactions are initiated mainly by the presence of intermediate bridging hydroxyl groups.

*Table 1. Calculated reaction energies and barriers for the H scission reactions for isolated species adsorbed to the surface.*

| Reaction | | Reaction energy (eV) | | Reaction barrier (eV) | |
| --- | --- | --- | --- | --- | --- |
| | | Minimum | Maximum | Minimum | Maximum |
| (1) | 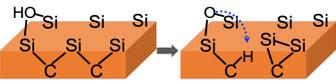 | -1.43 | -1.03 | 0.74 | 1.01 |
| (2) | 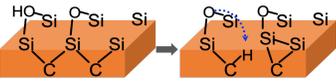 | -0.74 | -0.21 | 0.81 | 1.16 |
| (3) | 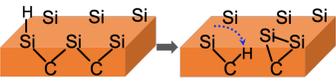 | 0.54 | 0.87 | 2.51 | 2.76 |
| (4) | 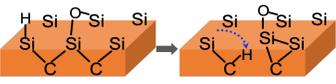 | 1.36 | 1.68 | 2.18 | 2.54 |
| (5) | 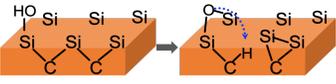 | -1.10 | -0.41 | 1.37 | 1.53 |

We note that apart from the hydrogen scission reaction, where H atoms attack a Si-C bond, there are other competing pathways for the release of hydrogen atom from the bridging hydroxyl group. One possibility is dissociation of the hydrogen atom from Si-OH-Si group to a surface Si atom to form a Si-H group (Si-OH-Si + *Si → Si-O-Si + Si-H). Hydrogen dissociation rate on the surface is highly dependent on the potential energy surface along the dissociation pathway. Our calculated activation energies for hydrogen dissociation are distributed over a wide range. Specifically, if no siloxanes were present on the dissociation pathway, the barriers would be relatively small, ranging from 0.15 eV to 0.23 eV. However, as siloxanes bridges already exist on the dissociation pathway, the barriers are significantly high, ranging from 1.13 eV to 1.26 eV (the higher concentration of the siloxane, the higher the barrier). This increase is due to H bonding with siloxanes on the way to find an unpassivated Si atom on the surface (Supplementary Sec. 5). Consequently, for the surface with a high concentration of siloxanes, it is expected that this dissociation mechanism is less competitive and a certain concentration of bridging hydroxyl groups could be stabilized on the surface. This stabilization effect can explain the observation of bridging hydroxyl groups in our MD simulations. Another pathway for releasing H from bridging hydroxyl groups is electrochemical dehydrogenation (Si-OH-Si → Si-O-Si + $H^+_{(aq)}$ + $e^-$) into the water environment. Such reaction is mostly dependent on the temperature, pH (controlling the chemical potential of proton in the solution), and electrode potential, $U$ (referenced to the Standard Hydrogen Electrode, SHE, and dependent of the material itself and water environment). Particularly, at neutral pH condition ($pH_{pzc}$~4.6-7.4 [37–39]), we found that for an electrode potential close to zero vs. SHE the dehydrogenation mechanism is energetically favorable, which may compete with the hydrogen scission reaction. As the electrode potential decreases, the dehydrogenation could be suppressed (Supplementary Fig. S6).

We have found H scission to be the necessary step in the water oxidation of SiC. We have focused the main discussion on Si-terminated surfaces, but for completeness we have performed similar calculations for C-terminated SiC surface and found the same qualitative conclusions (see Supplementary Sec. 7). In addition, although based on our current studies we cannot yet conclude if H scission is the rate limiting reaction in water attack on SiC, we can explore the question of what are possible reactions that follow H scission. With this question in mind, we perform additional *ab initio* MD simulations to simulate the oxidation process starting from surface configurations that have already undergone H scission reactions (see Reactions 1 and 2 in Table 1), where 20 $H_2O$ molecules are randomly inserted above the surface. We found that Si atoms that had been previously displaced above the surface and that as a result have excess electrons, are now easily attacked by $H_2O$ molecules leading to local oxidation, as shown in Fig. 3. Since the process in Reaction 1 after ~1 ps is similar to that in Reaction 2, here we only show the process in Reaction 2. Specifically, after ~10 ps, one more hydrogen scission reaction occurs on the surface, where a H atom from the neighboring hydroxyl group breaks the bond between the displaced Si atom and the sub-surface C atom. The displaced Si atom, with localized electron density highlighted by black arrow in Fig. 3(b), is further pushed away from the surface (to a distance of ~0.9 Å above the original position of this atom, see Fig. 3(b)). Subsequently, a new $H_2O$ molecule is dissociated into -H and -OH where the hydroxyl group forms Si-OH bond with the displaced Si atom and -H binds to a surface Si atom. The displaced Si atom (bonded to -OH) is further pulled away from the surface (to a distance of ~1.2 Å above the surface). At ~20 ps, this Si atom is again attacked by another $H_2O$ molecule, i.e., the molecule spontaneously dissociates around the displaced Si atom: the -OH group binds the displaced Si, creating a $Si(OH)_2$ group and the -H binds to a bridging siloxane group on the surface, forming a bridging hydroxyl group (see Fig. 3(c)). Over the next ~30 ps, the surface undergoes multiple deprotonation and hydrolysis reactions (Supplementary Fig. S7) and eventually forms a $SiO(OH)_2$ motif, see Fig. 3(d), and remains in this configuration for the rest of the simulation (up to 100 ps). As $H_2O$ molecules are adsorbed onto the surface, the water density in the solution is decreased, weakening the interaction between the solvent and the adsorbates in our simulations. In order to simulate the influence of solvent interaction on the stability of the $SiO(OH)_2$ motif, we then place additional five $H_2O$ molecules above this surface and we perform a new *ab initio* MD simulation for 10 ps. The $SiO(OH)_2$ motif remains stable in the presence of additional water molecules within our simulation time scales. Albeit the life-time of this motif cannot be determined from the short-time simulations we performed, the presence of the motif is consistent with observations from previous studies of dissolution of silicates[40–42]. In those studies, the $SiO(OH)_2$ group was argued to be a precursor for growth of defective silica in water, which is followed by dissolution[41,42]. We also performed an *ab initio* MD simulation for a surface that had not undergone hydrogen scission reaction, and we find that only the basic adsorption species are observed on the surface after up to 100 ps. These results are consistent with our DFT calculations of reaction energies and energy barriers (see Table 1), where we found that H scission from -H or -OH is energetically unfavorable. Our results suggest that perfect SiC surfaces are generally stable against oxidation in water and that attack of water on SiC happens through the H scission reaction. This reaction involves an intermediate species, which is the bridging hydroxyl group. The $SiO(OH)_2$ motif found in our simulations can be a pre-cursor to growth of either silica oxide scale or to the dissolution, based on what has been previously reported from studies of silica dissolution[40–42].

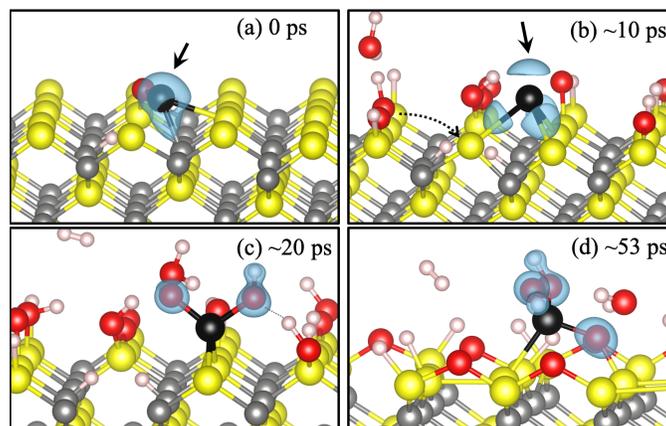

*Fig. 3. (a)-(d) Snapshots from an AIMD simulation after hydrogen scission reaction occurs on the surface. The corresponding charge density difference for those snapshots before and after bonding with the displaced Si atom. Blue regions depict the isocharge surface (at a value of 0.07 e/Å$^3$) of electron accumulation (positive). The black sphere represents the displaced Si atom after hydrogen scission reaction.*

## Conclusions

In summary, we have demonstrated that while in the most stable thermodynamic configuration H atoms passivate surfaces of SiC, H can also attack Si-C bonds through a H scission reaction. H scission occurs from intermediate metastable bridging hydroxyl groups that form on the surface of SiC in hydrothermal conditions. H scission makes it possible for additional water molecules to attack SiC, leading to formation of SiO(OH)$_2$ motifs. These motifs are known precursors to either silica formation or silica dissolution. The discovery of unexpected role of hydrogen in water environment provides new insight into mechanisms of wet oxidation of SiC for device processing and corrosion in hydrothermal environments encountered in nuclear reactors[11–13]. Hydrogen could potentially exhibit a similar behavior on surfaces of other semiconductors, where at present H formed during H$_2$O dissociation is considered to be a passivating species.

## Acknowledgment

Studies of Si-terminated surfaces reported in the main text were supported by DOE-BES grant #DE-FG02-08ER46493. To complete the study, we have carried out additional calculations of C-terminated surface under #DE-NE0008781.